\documentclass[aps,footinbib,twocolumn,showpacs,amsmath,amssymb,superscriptaddress,prb,groupedaddress]{revtex4}
\usepackage{graphicx,amsmath}
\usepackage{epstopdf}
\usepackage{amssymb}
\usepackage{grffile}
\usepackage[usenames]{color}
\usepackage{indentfirst}
\usepackage{float}
\usepackage{color}
\usepackage{mathrsfs}
\usepackage{dcolumn}
\usepackage{bm}
\usepackage[colorlinks=true,bookmarks=false,citecolor=blue,linkcolor=red,urlcolor=blue]{hyperref}
\usepackage{appendix}
\usepackage[T1]{fontenc}
\usepackage{lmodern}
\begin{document}
\title{Developments in the applications of density functional theory to fractional quantum Hall systems}
\author{Yi Yang}
\affiliation{Department of Physics and Chongqing Key Laboratory for Strongly Coupled Physics, Chongqing University, Chongqing 401331, People's Republic of China}
\author{Yayun Hu}
\affiliation{Zhejiang Lab, Hangzhou 311100, People's Republic of China}
\author{Zi-Xiang Hu}
\email{zxhu@cqu.edu.cn}
\affiliation{Department of Physics and Chongqing Key Laboratory for Strongly Coupled Physics, Chongqing University, Chongqing 401331, People's Republic of China}

\begin{abstract}
The fractional quantum Hall effect remains a captivating area in condensed matter physics, characterized by strongly correlated topological order, which manifests as fractionalized excitations and anyonic statistics. Numerical simulations, such as exact diagonalization, density matrix renormalization group, matrix product states, and Monte Carlo methods, are essential to examine the properties of strongly correlated systems. Recently, density functional theory has been employed in this field within the framework of composite fermion theory. This paper systematically evaluates how density functional theory approaches have addressed fundamental challenges in fractional quantum Hall systems, including ground state and low-energy excitations. Special attention is given to the insights provided by density functional theory regarding composite fermion behavior, edge effects, and the nature of fractional charge and magnetoroton excitations. The discussion critically examines both the advantages and limitations of these approaches, while highlighting the productive interplay between numerical simulations and theoretical models. Future directions are explored, particularly the promising potential of time-dependent density functional theory for modeling non-equilibrium dynamics in quantum Hall systems.
\end{abstract}
\date{\today} 
\pacs{73.43.Lp, 71.10.Pm}
\maketitle
\section{Introduction}
The fractional quantum Hall effect (FQHE), discovered in 1982~\cite{DCTHLSACG}, stands as one of the most profound discoveries in condensed matter physics. Emerging just two years after the integer quantum Hall effect, the FQHE unveiled a highly complex world of two-dimensional electron systems in high magnetic fields. Unlike its integer counterpart, it defies conventional explanations based on single-particle physics. Instead, it heralds the emergence of exotic quantum states governed by intricate many-body interactions, where electrons collectively organize into topological phases. This discovery not only reshaped our understanding of quantum matter, but also introduced concepts such as topological order, fractional charge, and statistics~\cite{RBL, topo1, topo2, topo3}, ideas that continue to challenge and inspire theoretical and experimental pursuits.

Central to the FQHE is its strongly correlated nature. Early theoretical breakthroughs, such as Laughlin's seminal wavefunction~\cite{RBL}, provided a visionary framework for fractionally charged quasiparticles. Yet, the complexity of these systems-where conventional perturbative methods fail-demands nontrivial theoretical tools and computational approaches. Traditionally, numerical simulations have been crucial in shedding light on the complexities of FQHE. From exact diagonalization of small systems to large-scale Monte Carlo studies, computational techniques have validated theoretical models, predicted new phases, and bridged the gap between microscopic interactions and macroscopic observables~\cite{numerical1,gsenergy2,numerical2,numerical3,numerical4,numerical5,numerical6}. For example, numerical diagonalization has been crucial in confirming the predictions of the Laughlin wavefunction regarding fractional statistics and incompressible quantum Hall states by accurately solving many-body Hamiltonians in finite systems.  However, the exponential growth of Hilbert space with particle numbers and the challenges of incorporating disorder, realistic long-range interaction, and real material complexities have highlighted the need for efficient and scalable computational frameworks.

It is within this context that density functional theory (DFT) enters as a compelling, albeit unconventional, candidate for studying the FQHE. Traditionally, DFT has been a cornerstone of electronic structure calculations for weakly correlated systems, leveraging the Hohenberg-Kohn theorems to map interacting many-body problems into effective single-particle equations~\cite{HKDFT1,HKDFT2,HKDFT3,HKDFT4,HKDFT5,HKDFT6,HKDFT7,HKDFT8,HKDFT9}. While standard DFT has been successful for weakly correlated systems, its application to strongly correlated FQHE was initially met with skepticism due to limitations in capturing topological order. However, recent advances have begun to challenge this narrative. The proposals for density functional approaches, which incorporate the composite fermion (CF) theory~\cite{Jain,Jainbook}, aim to encode topological order and fractional statistics within a DFT framework. These developments, combined with the method's inherent ability to handle realistic experimental conditions such as edge potentials, suggest a promising synergy between the computational efficiency of DFT and the rich physics of FQHE.

The application of DFT to FQHE began with early studies by Ferconi \textit{et al.}~\cite{DFT1}, who explored the modeling of the edge profiles of FQH liquids under different confining potentials. These studies revealed incompressible stripes at smooth edge potentials and contributed to a deeper understanding of the spatial variations of edge states. Heinonen \textit{et al.}~\cite{DFT2} further extended the application of DFT to FQH systems by introducing ensemble DFT, solving the Kohn-Sham (KS) equations, and providing initial analyses of the edge structure and edge reconstruction in certain FQH states. These pioneering works laid the groundwork for the use of DFT in the FQHE field. As CF theory developed, Zhang \textit{et al.}~\cite{DFT3} developed a robust and comprehensive current DFT framework to assess the ground state of FQHE. Zhao \textit{et al.}~\cite{DFT4} illustrated the effectiveness of DFT in modeling CF excitations, enhancing the exchange-correlation energy of CFs, and employing the Thomas-Fermi approximation to represent the kinetic energy of CF. This approach improved the accuracy of edge physics~\cite{edge1,edge2,edge3,topo2} and provided a more precise mapping of the real electron system to an auxiliary CF system, solving the KS equations to predict the physical properties of the actual system. These advances revealed the hierarchical nature of edge reconstruction at low temperatures and large separations between the background layer and the electronic layer. Hu \textit{et al.}~\cite{DFT5} made a significant progress by using magnetic field DFT in the KS scheme to explore the properties of FQHE. A precise calculation of the topological properties including the fractional charge and statistics have been achieved for the first time using DFT in the strongly correlated systems. Subsequent research expanded this framework to investigate Abelian anyons and discrete crystal FQH states~\cite{DFT6, DFT7}, demonstrating the phase transition from liquid to crystal phases influenced by different exchange-correlation potential strengths. Recently, we~\cite{DFT8} utilized DFT to explore magnetoroton excitation and its chirality, uncovering the interaction between edge excitations and bulk correlations by examining CF excitons.

This work reviews recent advances in DFT applications to the FQHE, covering ground state, anyon excitations, statistics, magnetoroton excitations, and the topological properties. A particular focus is placed on the comparative advantages of DFT approaches over traditional methods like Monte Carlo (MC) simulations. The discussion extends to emerging research directions, including the promising application of DFT to dynamic properties in FQHE systems. The paper is organized as follows. Sec.~\ref{sec2} presents the CF-based DFT framework for FQHE studies. Sec.~\ref{sec3} analyzes DFT investigations of ground state density profiles and energy convergence. Sec.~\ref{sec4} surveys progress in characterizing fractional charge excitations and quasiparticle statistics, including topological spin calculations. Sec.~\ref{sec5} explores DFT treatments of bulk collective excitations and magnetoroton dispersion relations. Finally, Section~\ref{sec6} summarizes key findings and discusses future prospects for DFT in studying dynamic phenomena in FQH systems.

\section{CF-based DFT Method in FQHE}\label{sec2}
\textit{CF theory.} The CF picture is a convenient way to understand various aspects of FQHE~\cite{Jain}. A composite particle is obtained by attaching $m$ quanta of the Chern-Simons flux to an electron: It is a composite boson when $m = 2p + 1$ is odd, and a composite fermion when $m = 2p $ is even. The system of electrons in the magnetic field turns out to be the system of composite particles in the magnetic flux and Chern-Simons flux, where composite particles experience a residual effective magnetic field.

\begin{figure}[ht]        
        \centering
        \includegraphics[width=8cm,height=2cm]{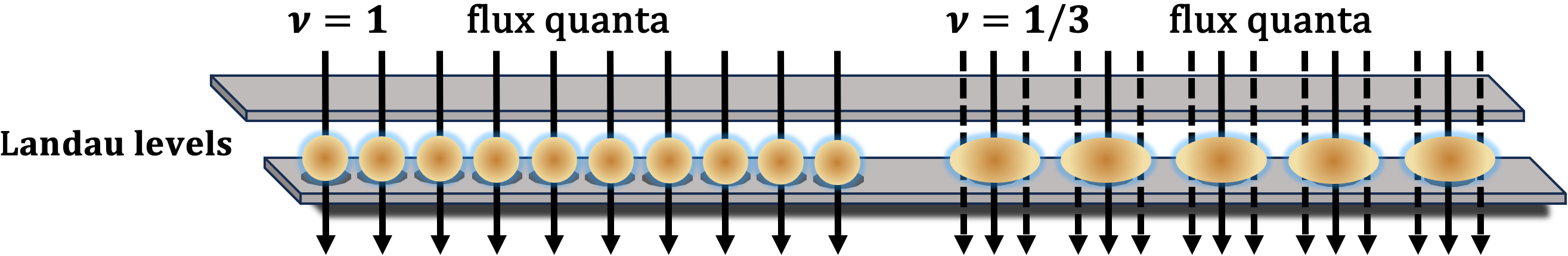}        
        \caption{\label{modelcf} Schematic illustration of the mapping between $\nu = 1/3$ FQH state and $\nu = 1$ IQH state after binding two Chern-Simons flux to each electron in CF theory.}
\end{figure}

In Jain's sequence at $\nu=n/(2n \pm 1)$, each electron combines with two flux quanta to form a CF, which is subject to a diminished effective magnetic field given by $B^*=(1-2\nu)B$. The associated effective magnetic length can be expressed as $l^*=\sqrt{\hbar c/eB^*}$. The effective filling factor of composite fermions is $\nu_{eff} = (2n+1)\nu = 2n$. For the well-known Laughlin state at $\nu=1/3$, as schematically illustrated in Fig.~\ref{modelcf}, the associated CF state behaves as a weakly interacting IQH state which is suitable for DFT algorithm. The low-energy physics, which encompasses the ground states, charge excitations, and neutral excitations, can be represented by CFs filling $\Lambda$ levels ($\Lambda$Ls) within this effective magnetic field~\cite{Jainbook}. For instance, creating a quasihole involves vacating an orbital within the filled $\Lambda$Ls, whereas a quasiparticle is generated by filling an orbital in the unoccupied $\Lambda$Ls. Moreover, a CF exciton consists of an excitation involving a quasiparticle-quasihole pair in two adjacent $\Lambda$Ls, which will be elaborated on below.
 
\textit{KS Equations.} The DFT framework within CF theory is formulated through the solution of KS equations for CFs. Following the approach developed in Ref.~\onlinecite{DFT5}, the KS equations take the form:
\begin{equation}
\label{ksequation1}
	\left[T^*+V^*_{\rm KS}(\bm{r}) \right] \psi_{\alpha}(\bm{r}) = \epsilon_{\alpha} \psi_{\alpha}(\bm{r})
\end{equation}
where the parameter $\alpha$ represents the orbital index in each $\Lambda$Ls.  In the symmetric gauge, the CF orbitals maintain rotational symmetry, it could be the angular momentum $m$ and the energy level $n$ ($\Lambda$ level) of the single-particle orbital. Therefore, the KS orbits are labeled by the index $(n,m)$. The KS orbital wavefunctions are written as:
\begin{equation} \label{singleorbital}
	\psi_\alpha(\bm{r})=\frac{R_\alpha(r)}{\sqrt{2\pi r}}\mathrm{e}^{-\mathrm{i}m_\alpha \theta}
\end{equation}
where $R_\alpha(r)$ is the radial part, and $m_\alpha$ is the angular momentum quantum number of CF orbital. Each KS equation in Eq.~(\ref{ksequation1}) is solved using the finite-difference method for nonzero $m_\alpha$. The explicit forms of $T^*$ and $V_T^*$ are derived in next sections. For $m_\alpha=0$, a basis expansion method is used to handle the singularity at $r=0$, as will be discussed later.  With these KS orbital wavefunctions,  the CF density is computed as 
\begin{equation}
	\rho(\bm{r}) = \sum_{\alpha} c_\alpha |\psi_\alpha(\bm{r})|^2
\end{equation}
where $c_\alpha=1$ for fully occupied orbitals at zero temperature, which guarantees that the total number of electrons is $\sum_\alpha c_\alpha=N_e$. 

In Eq.~(\ref{ksequation1}), the structure of the KS potential is represented as $V^*_{\rm KS}(\bm{r})=V_{\rm H}(\bm{r})+V_{\rm ext}(\bm{r})+V_{\rm xc}^{*}(\bm{r})+V^*_{\rm T}(\bm{r})$. Here, $V_{\rm H}, V_{\rm ext}, V_{\rm xc}^{*}, V^*_{\rm T}$ are the Hartree, external, exchange-correlation, and kinetic energy potentials, respectively.  The detailed expressions of KS potentials are as follows
\begin{eqnarray}\label{ksequation2}
  	T^* &=& \frac{1}{2m^*}\left(\bm{p}+\frac{e}{c}\bm{A}^*(\bm{r})\right)^2 \nonumber\\
		V_{\rm H}(\bm{r})&=&\int d\bm{r}'\frac{\rho(\bm{r}')}{|\bm{r}-\bm{r}'|} \nonumber\\
		V_{\rm ext}(\bm{r}) &=& \left\{ 
		\begin{array}{ll}
        -\sqrt{\frac{2}{3}N_e} \frac{2E(\frac{r^2}{6N_e})}{\pi} ,\frac{r}{\sqrt{6N_e}}\leq  1 \nonumber\\     
        -\sqrt{\frac{2}{3}N_e} \frac{_2F_1(\frac{1}{2},\frac{1}{2};2;\frac{6N_e}{r^2})}{2r/\sqrt{6N_e}} ,\frac{r}{\sqrt{6N_e}}\geq 1 \\
        \end{array} \right.  \nonumber\\ 
		V_{\rm xc}^*(\bm{r})&=&\frac{3}{2}a[2\pi l^2 \rho(\bm{r})]^{\frac{1}{2}}+(2b-f)[2\pi l^2 \rho(\bm{r})]+g \nonumber\\
		V^*_{\rm T}(\bm{r})&=&\sum_{\alpha}c_{\alpha}\langle\psi_{\alpha}|\frac{\delta T^*}{\delta \rho(\bm{r})}|\psi_{\alpha}\rangle  
\end{eqnarray}
where the parameters $a = -0.78213, b = 0.2774, f = 0.33, g = -0.04981$ originate from the local density approximation (LDA) for the exchange-correlation energy. The effective vector potential is expressed as $\bm{A}^*(r) = \frac{1}{r} \int_0^r r' B^*(r') dr' \bm{e}_{\theta}$, with $\nabla \times \bm{A}^*(r) = B^*(r)\bm{e}_{\rm z} = \left[B - 2\rho(r)\phi_0\right] \bm{e}_{\rm z}$. Here, $\rho_{\text{b}} = \nu_0/2\pi l^2$ represents the uniform background charge density on a disk of radius $R_{\text{b}}$, where the condition $\pi R_{\text{b}}^2 \rho_{\text{b}} = N_e$ is satisfied.  An experimental sample consisting of a modulation-doped GaAs/GaAlAs heterostructure ~\cite{Pfeiffer89} can be described as a two-dimensional electron layer at the GaAs and GaAlAs interface along with a uniform positive background charge originating from remote dopants at a distance $d$. The assumption of $d=0$ has been widely adopted to ensure maximum confinement and circumvent edge reconstruction. The external potential $V_{\rm ext}(\bm{r})$ is calculated as described in Ref.~\onlinecite{backpotential}. In polar coordinates, the Hamiltonian $\mathcal H^*=T^*+V^*_{\rm KS}(r)$ in Eq.~(\ref{ksequation1}) and Eq.~(\ref{ksequation2}) is:
\begin{eqnarray}\label{ksStarpolarform}
    \mathcal H^*&=&\frac{1}{2m^*}\bigg[-\hbar^2\big(\frac{\partial^2}{\partial r^2}+\frac{1}{r^2}\frac{\partial^2}{\partial \theta^2} \nonumber \\
    &+&\frac{1}{r}\frac{\partial}{\partial r}\big)+\frac{\hbar e}{c}\mathcal{B}(r)\frac{\hat{L}_z}{\hbar}+\frac{e^2}{c^2}\frac{r^2\mathcal{B}^2(r)}{4}\bigg]+V^*_{\rm KS}
\end{eqnarray}
where $\mathcal{B}(r)=\frac{1}{\pi r^2}\int_0^r 2\pi r' B^*(r')dr'$. Utilizing Eq.~(\ref{ksequation1}), Eq.~(\ref{singleorbital}), and Eq.~(\ref{ksStarpolarform}), in conjunction with the eigenvalue equation $\mathcal H^* = T^*+V^*_{\rm KS}(\bm{r}) $ and the condition $\frac{1}{2}\hbar \omega_B l^2=\frac{\hbar^2}{2m^*}$, with the assumption $l^2=\frac{\hbar c}{eB}=1$, the KS equations take the following form:
\begin{eqnarray}
		&&\frac{1}{2}\hbar \omega_B [-\frac{\partial^2 }{\partial r^2}+\frac{m_\alpha^2-1/4}{r^2}+\frac{r^2 \mathcal B^2(r)}{4B^2}\nonumber \\
        &-&m_\alpha \frac{\mathcal B(r)}{B}]{R}_{\alpha}(r)+V^*_{\rm KS}{R}_{\alpha}(r)=\epsilon_\alpha {R}_{\alpha}(r) 
\end{eqnarray}
where the angular momentum $\hat{L}_z=-\mathrm{i}\hbar \frac{\partial }{\partial \theta}$ and $\langle L_z \rangle = -m_\alpha \hbar$. It is crucial to note that $\langle L_z \rangle = \pm m_\alpha \hbar$ is contingent upon how the wave function is defined, specifically $\psi_{\alpha}(\bm{r})=\frac{{R}_{\alpha}(r)}{\sqrt{2\pi r}}\exp(\pm im_{\alpha}\theta)$. The radial wave function ${R}_{\alpha}$ satisfies the equation:
\begin{equation}\label{ksStar1dreduced}
  \tilde{\mathcal H}^*_{m_\alpha}(r){R}_{\alpha}(r)=\epsilon_{\alpha}{R}_{\alpha}(r)
\end{equation}
where the Hamiltonian is defined as:
\begin{eqnarray} \label{1DksStarpolarform} 
	& &\tilde{\mathcal H}^*_{m_\alpha}(r)=V^*_{\rm KS}+T^* = V^*_{\rm KS}   \\
         &+&\frac{1}{2} \hbar \omega_B [-\frac{\partial^2}{\partial r^2}+\frac{m_\alpha^2-\frac{1}{4}}{r^2}-m_\alpha\frac{\mathcal{B}(r)}{B}+\frac{\mathcal{B}^2(r)}{B^2}\frac{r^2}{4}] \nonumber
\end{eqnarray}

The potential $V_T^*$ is obtained through the following derivation:
\begin{eqnarray}\label{DeltaTStarDef}
	&&V^*_{\rm T}(\bm{r})=\sum_{\alpha}c_{\alpha}\langle\psi_{\alpha}|\frac{\delta T^*}{\delta \rho(\bm{r})}|\psi_{\alpha}\rangle \nonumber \\
    &=&\frac{1}{2}\hbar w_B \sum_{\alpha}c_{\alpha}\int d r' R_\alpha(r') \left( \frac{4m_\alpha  l_B^2  }{r'^2}- 2\frac{\mathcal{B}(r')}{ B } \right) \nonumber \\
    &&\theta(r'-r) R_\alpha(r')  
\end{eqnarray} 
 where the variational derivative is:
 \begin{eqnarray} 
		\frac{\partial A^*(r)}{\partial \rho(r'')}&=&-\frac{\phi_0}{\pi r }\theta(r-r'')  
\end{eqnarray}
 in which the $\theta(r)$ is step function. Eq.~(\ref{1DksStarpolarform}) can be numerically solved for every non-zero angular momentum $m_\alpha$ utilizing the finite-difference technique.

For $m_{\alpha}=0$, a basis expansion method is typically adopted to handle the singularity at $r=0$. The matrix form of $\tilde{\mathcal H}_{m_0}^*$ can be obtained using the basis set $\textbf{H}_0=\{\eta_{n,m=0}({\bm{r}},{\mathcal{B}_0}), n=0,1,\ldots,N_{\rm L}\}$ in the angular momentum $m=0$ subspace, where the basis $\eta_{n,m}({\bm{r}},{\mathcal{B}_0})$ are Landau level (LL) wave functions with symmetry gauge and effective magnetic length $l_{\mathcal{B}_0}$, and $N_{\rm L}$ is the cutoff for $\Lambda$L, which is fixed at $N_{\rm L}=30$ in this study. 

The determination of $l_{\mathcal{B}_0}$ is described as follows. In the $\textbf{H}_0$ subspace, the matrix elements $\mathcal{H}^*_{(n',n)}$ take the form:
\begin{eqnarray} 
		\mathcal{H}^*_{(n',n)}&=&T^{*}_{(n',n)}+V^{*}_{\text{KS}{(n',n)}}\nonumber\\
 	T^{*}_{(n',n)}&=&\hbar \omega_{\mathcal{B}_0}(n+\frac{1}{2})\delta_{n'n}+\langle\eta_{n',0}({\bm{r}},{\mathcal{B}_0})|V_0|\eta_{n,0}({\bm{r}},{\mathcal{B}_0})\rangle \nonumber\\
	V^{*}_{{\textrm{KS}}{(n',n)}}&=&\langle\eta_{n',0}({\bm{r}},{\mathcal{B}_0})|V^{*}_{\rm{KS}}|\eta_{n,0}({\bm{r}},{\mathcal{B}_0})\rangle
\end{eqnarray}
where
\begin{equation}
	\begin{split}
        V_0 &= \frac{1}{8}\hbar \omega_{\mathcal{B}_0} \frac{\mathcal{B}^2({r})-\mathcal{B}^2_{0}}{\mathcal{B}^2_{0}}(r/l_{\mathcal{B}_0})^2
	\end{split}
\end{equation}
where $\omega_{\mathcal{B}_0}=\frac{e\mathcal{B}_{0}}{m^*c}$. Eigenfunctions $c_{n',n}$ and eigenvalues $\epsilon_{n,0}$ emerge from the diagonalization of $\mathcal{H}^*_{(n',n)}$. The single CF orbital with $m_\alpha=0$ can then be expressed as a linear superposition of these eigenfunctions:
\begin{equation}
   \psi_{n,0}(\bm{r})=\sum_{n'=0}^{N_{\rm L}} c_{n',n}\eta_{n',0}(\bm{r}, {\mathcal{B}_0})
\end{equation}
a reasonable choice for $\mathcal{B}_0$ is given by:
\begin{equation}\label{m0approxiamtion}
\hbar\frac{e\mathcal{B}_{0}}{m^*c} =\frac{\Delta_{m_\alpha=-1}+\Delta_{m_\alpha=1}}{2}
\end{equation}
where $\Delta_{m_\alpha=\pm1}$ represents the average energy gap between the lowest two energy levels in the $m_\alpha=\pm 1$ subspace, as obtained using the finite difference method. This approximation is valid because orbitals with $m_\alpha=0$ overlap spatially with those of $m_\alpha=\pm1$ and therefore experience a similar effective magnetic field. 

The KS equations presented in Eq.~(\ref{ksequation1}) are resolved self-consistently using the standard KS-DFT iterative approach by integrating the aforementioned methods. The procedure is outlined as follows:
\begin{itemize}
  \item \textbf{Initialization:} Start with an initial electron density $\rho_{\text{in}}$, which is typically considered as the background density.
  \item \textbf{Calculation:} First, determine the kinetic energy $T^*$ and the effective potential $V^*_{\rm KS}(\bm{r})$, then solve the KS equations to find the orbitals $\psi_\alpha(\bm{r})$. The output density $\rho_{\rm out}$ is subsequently derived from these orbitals.
  \item \textbf{Convergence Check:} Calculate the relative difference $\Delta = \frac{\int d\bm{r} |\rho_{\rm out} - \rho_{\rm in}|}{N}$. The solution is deemed to be converged if $\Delta < 10^{-5}$.
  \item \textbf{Update:} If convergence is not achieved, update the input density using a mixing scheme: $\rho'_{\rm in} = \lambda \rho_{\rm in}+(1-\lambda)\rho_{\rm out}$, with $\lambda= 0.95$. Repeat until convergence is reached.
\end{itemize}

Consequently, the total energy of the system consists of various components:
\begin{equation}
   E=E_{\text{xc}}^*+E_{\text{T}}^*+E_{\text{H}}+E_{\text{ext}}+E_{\text{bb}}
\end{equation}
where
\begin{equation}
	\begin{split}
		&E_{\text{xc}}^*=\int \large[a(2\pi)^{1/2}\rho^{3/2}+(2b-f)\pi\rho^2+g\rho \large] d\bm{r},\\
		&E_{\text{T}}^*= \sum_\alpha \langle \psi_\alpha|T^*|\psi_\alpha\rangle,\\
		&E_{\text{H}} =  \frac{1}{2} \int d\bm{r} \int d\bm{r}'\frac{\rho(\bm{r}')\rho(\bm{r})}{|\bm{r}-\bm{r}'|},\\
		&E_{\text{ext}}= \int V_{\rm ext} \rho(\bm{r}) d\bm{r},\\
		&E_{\text{bb}}= \frac{8N_e}{3\pi}\sqrt{\frac{N_e}{6}}.
	\end{split}
\end{equation}
In this context, $E_{\text{H}}$ denotes the electron-electron interaction, $E_{\text{ext}}$ means the interaction with the external potential, including background potential, while $E_{\text{bb}}$ is a constant associated with the background-background interaction. The term $E_{\text{T}}^*$ refers to the kinetic energy of the CFs and $E_{\text{xc}}^*$ represents the exchange correlation energy, adjusted for the discrepancies between the auxiliary and real systems.

For a homogeneous ground state, only $E_{\text{xc}}^*$ contributes in the thermodynamic limit, as all other terms cancel out:
\begin{equation}
	\begin{split}
		\lim_{N_e\rightarrow \infty}E/N_e&=E_{\mathrm{xc}}/N_e \\
		&=\frac{\int \large[a(2\pi)^{1/2}\rho^{3/2}+b(2\pi)\rho^2+g\rho \large] d\bm{r}}{N_e}
	\end{split}
\end{equation}

\section{Ground state}\label{sec3}
As a prototypical example, the $\nu = 1/3$ Jain state ($n = 1$) serves as an important test case. This system is exactly described by the Laughlin wavefunction~\cite{RBL}
\begin{equation}\label{Lauwf}
\Psi_{\text{L}} = \prod_{i<j}(z_i - z_j)^3 \mathrm{e}^{-\sum_i |z_i|^2/4l_B^2}
\end{equation}
precisely characterizes the ground-state properties. It represents an exact zero energy eigen state for a short-range interaction characterized by Haldane pseudopotential $V_1$\cite{haldanepps}. The model wave function could be an input to the MC simulation of the Laughlin state~\cite{backpotential,yangyi} to explore its properties. 

\begin{figure}[ht]        
\center{\includegraphics[width=8cm]  {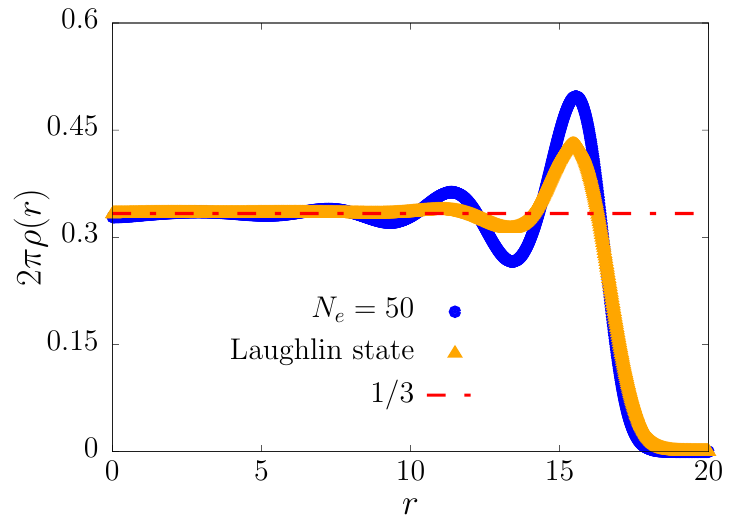}}        
\caption{\label{density} Density profile of the $\nu=1/3$ FQH state: DFT ground state vs. Laughlin state. The edge density oscillations are more pronounced in DFT due to Coulomb interactions.}
\end{figure}

In Fig.~\ref{density}, the radial electron density for the ground state is calculated for $\nu = 1/3$ in a system comprised of $50$ electrons. Per the CF mapping, the ground state is established by occupying the lowest $\Lambda$ level with the initial $N_e$ CF orbitals. Additionally, the calculated density profile is compared with results obtained from MC simulations of the Laughlin wavefunction in Eq.~(\ref{Lauwf}). Clearly, integrating the Coulomb interactions within DFT leads to more pronounced oscillations at the edge, which demonstrates a long-range correlation. Within the bulk, the density approaches $1/3$, indicating a uniform quantum liquid.

\begin{figure}[ht]        
\center{\includegraphics[width=8cm]  {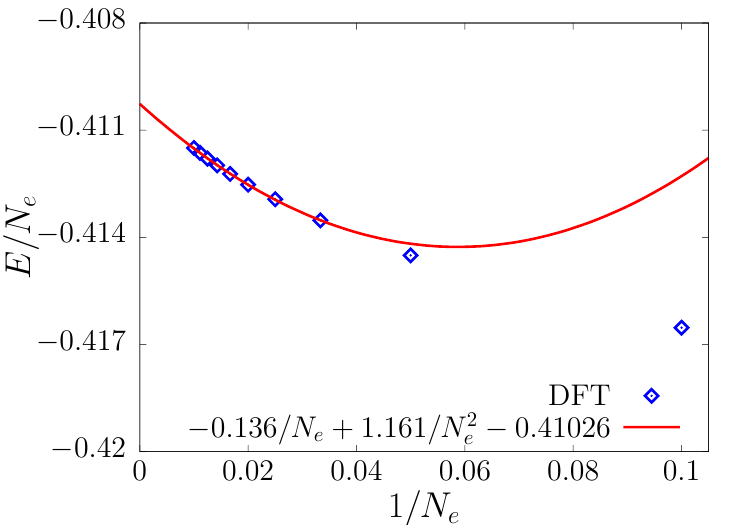}}        
\caption{\label{dft_gs_energy} The ground state energy per CF at $\nu=1/3$. The extrapolated value in thermodynamic limit obtained ($-0.41026 \pm 0.00003$) agrees well with previous reported value $\epsilon \approx -0.41015$~\cite{backpotential,gsenergy2,gsenergy1}. } 
\end{figure}
After achieving convergence in the ground states, the energy for various system sizes is calculated and extrapolated to the thermodynamic limit. These findings are shown in Fig.~\ref{dft_gs_energy}. Extrapolation of data from larger systems ($N_e \geq 30$) yields an energy per particle of $-0.41026 \pm 0.00003$ within the DFT framework, in excellent agreement with the previously reported value $\epsilon \approx -0.41015$ from other methods~\cite{backpotential,gsenergy2,gsenergy1}.

\section{quasi-particle excitation} \label{sec4}
With the ground state properties established, DFT can be employed to calculate the fractional charge and statistics of the quasiparticle/quasihole excitations, showcasing its effectiveness in accurately capturing more topological characteristics of FQH states.

\begin{figure}[ht]        
\center{\includegraphics[width=8cm]  {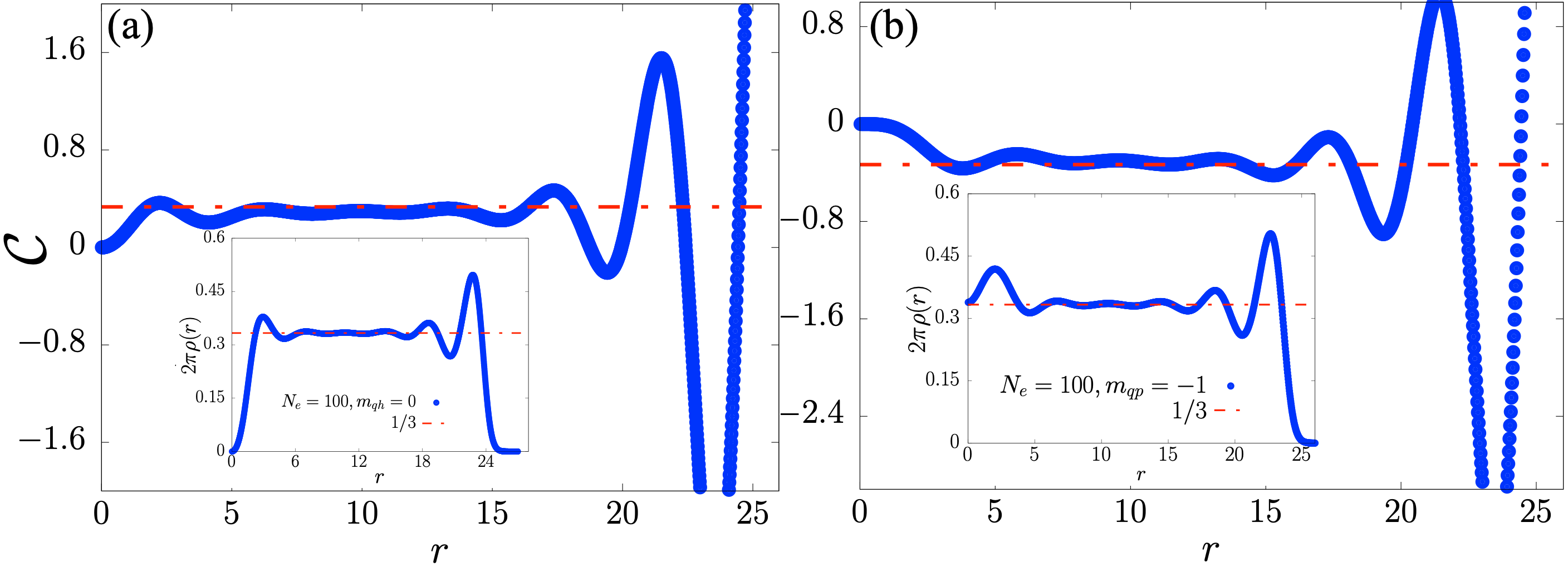}}        
\caption{\label{cumulation} Charge accumulation for (a) a quasihole and (b) a quasiparticle at the origin in the $\nu = 1/3$ FQH state. (a) Quasihole with charge $e/3$ at position $(n=0,~m=0)$, and (b) Quasiparticle with charge $-e/3$ at position $(n=1,~m=-1)$. The inset shows the corresponding electron density distribution for each case.} 
\end{figure}

\textit{Fractional Charge excitation.}
Within CF theory, quasiholes and quasiparticles are formed by vacating or filling KS orbitals. A quasihole arises at the disk's center by vacating the $0\Lambda L$ angular momentum orbital where $m=0$. In contrast, a quasiparticle is produced when the CF fills the $1\Lambda L$ orbit with $m=-1$. The charge can be evaluated through computation of the charge accumulation $\mathcal{C}$:
\begin{equation}
	\mathcal{C}(r) = \int_0^r [\frac{\nu}{2\pi}-\rho(r')] d^2r'
\end{equation}
Fig.~\ref{cumulation} illustrates the charge accumulation profiles for a quasihole and a quasiparticle situated at the disk's center in the $\nu = 1/3$ FQH state. In panel (a), the quasihole is located at $(n=0,~m=0)$, while in panel (b), the quasiparticle is at $(n=1,~m=-1)$. The insets provide the respective electron density distributions for each scenario. In both cases, the fractional charges $\mathcal{C}$ approach $\pm e/3$ in the asymptotic limit, i.e., far away from the center in the bulk ($r \sim 10l_B$).

\begin{figure}      
\center{\includegraphics[width=8cm]  {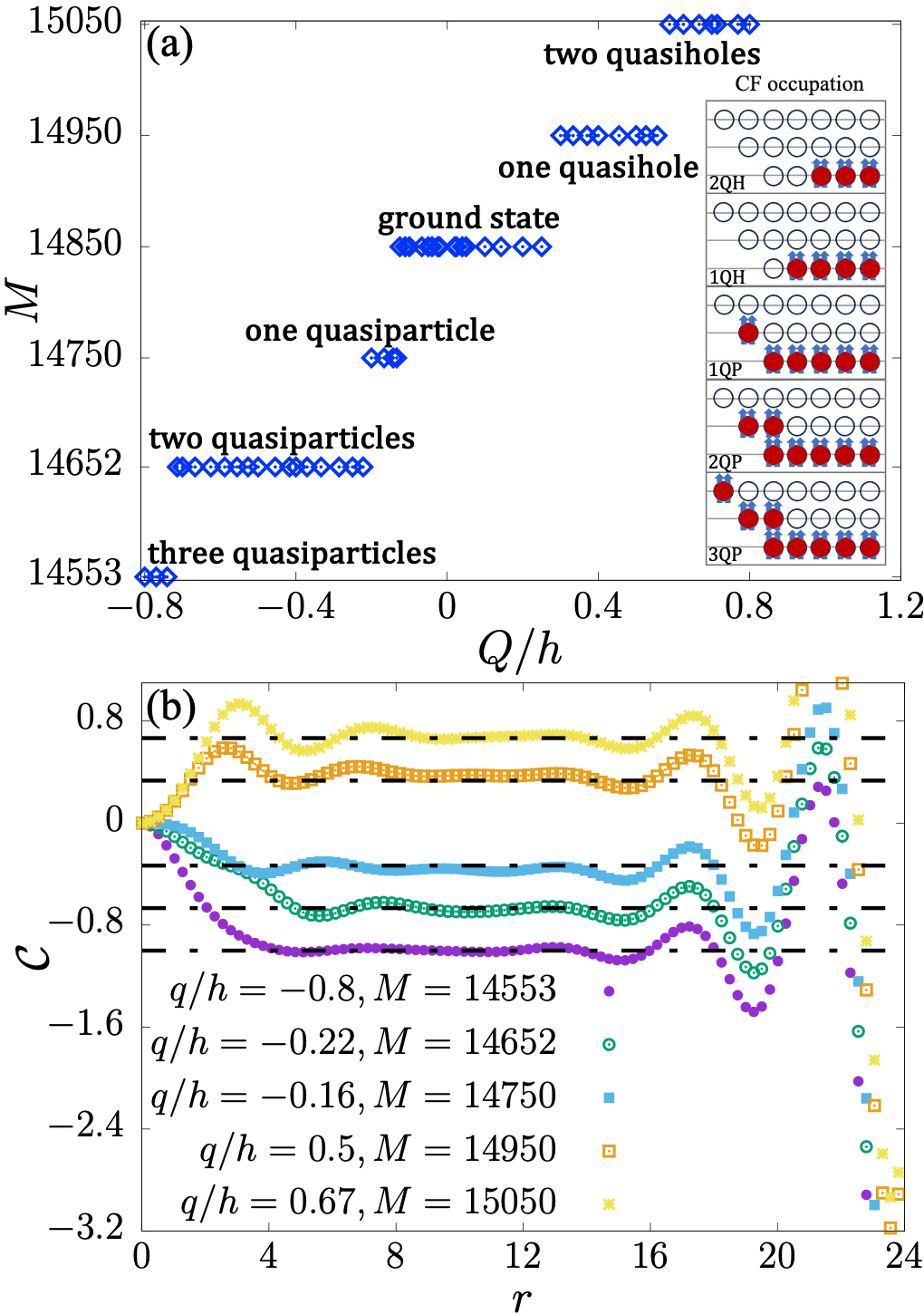}}        
\caption{\label{platue} (a) The total angular momentum of the global ground state as tuning the magnitude of $Q/h$  for 100 electrons at $\nu = 1/3$. The inset schematic diagram shows the CF occupations. (b) The charge accumulation $\mathcal{C}$ for all kinds of excitations.} 
\end{figure}

The theoretical framework effectively models a scanning tunneling microscope (STM) tip positioned at the disk center to generate quasiparticle or quasihole excitations. In this approach, the tip is represented as a charged impurity with $Q=\pm e$ located at height $h$ above the disk center, producing an electrostatic potential of the form:
\begin{equation}
	V_{\text{imp}}(r)=\frac{Q}{\sqrt{|r|^2+h^2}}
\end{equation}
The height $h$ serves to modify the impurity's strength. By adjusting $h$ and the sign of the charge $Q$, it is possible to create one or several quasiparticles and quasiholes, each with a charge that is an integer multiple of $1/3$. These excitations could be identified by the total angular momentum $M$ of the global ground state. In a system containing $N_e$ electrons, the ground state is located within a subspace where $M = 3N_e(N_e-1)/2$. The $pe/3$ charged quasihole state is realized by shifting all the CFs $p$ KS orbits and thus experiences a momentum shift $\Delta M = pN_e$ greater than the ground state. While the $-pe/3$ charged quasiparticle state is realized by moving $p$ CFs to the innermost orbital in higher $\Lambda$Ls as shown in the inserted schematic diagram of Fig.~\ref{platue}(a)~\cite{DFT5}, thus experiences a momentum shift $\Delta M = -\sum_{i=1}^p (N_e -i) + m_p$ smaller than the ground state where $m_p$ is the total momentum of CFs in higher $\Lambda$ Ls, i.e., $m_p = -1$ for $p = 1, 2$ and $m_p = -3$ for $p=3$. The total angular momentum of the global ground state for 100 electrons is shown in Fig.~\ref{platue}(a) under varying impurity strength $Q/h$~\cite{hu08prb}. In this situation, the one quasihole and two quasiholes states are characterized by quantum numbers $M = 14950$ and $M = 15050$, while the one quasiparticle, two and three quasiparticles states have quantum numbers $M = 14750$, $M = 14652$, and $M = 14553$. Fig.~\ref{platue}(b) shows the charge accumulation $\mathcal{C}$ for all excitations, demonstrating that each carries a charge of integer multiples of $e/3$.

\textit{Fractional Statistics.}
For systems with one- and two-quasihole states, fractional statistics can be verified through computation of the Berry phase during braiding processes. When one quasihole is fixed at the disk's center, braiding is equivalent to adiabatically rotating the system. The Berry phase in this context is defined as
\begin{equation}
\phi_B(\theta) = \mathrm{i} \oint_{\theta} \langle \Psi(\theta)|\partial_{\theta}|\Psi(\theta)\rangle.
\end{equation} 
Alternatively, the rotation might be produced by the angular momentum component $\hat{L}_z$ as follows:
\begin{eqnarray}
|\Psi(\theta + \delta \theta)\rangle &=& \exp(-\mathrm{i}\hat{L}_z\delta\theta/\hbar) |\Psi(\theta)\rangle \nonumber \\
&\simeq& (1-\mathrm{i}\hat{L}_z\delta\theta/\hbar) |\Psi(\theta)\rangle.
\end{eqnarray} 
These results demonstrate that
\begin{equation}
\phi_B(\theta) = \frac{1}{\hbar} \oint_{\theta} \langle \Psi(\theta)|\hat{L}_z|\Psi(\theta)\rangle  = \frac{2\pi}{\hbar} \langle \hat{L}_z \rangle.
\end{equation}
With exact wavefunctions, the braiding statistical phase of two quasiholes can be expressed by the Berry phase of the two-quasihole state subtracting the Berry phases for the two single-quasihole states as shown in Ref.~\onlinecite{Lijie}. 
\begin{widetext}
\begin{figure}[ht]        
    \centering
    \begin{minipage}{\textwidth}
        \centering
        \includegraphics[width=16cm]{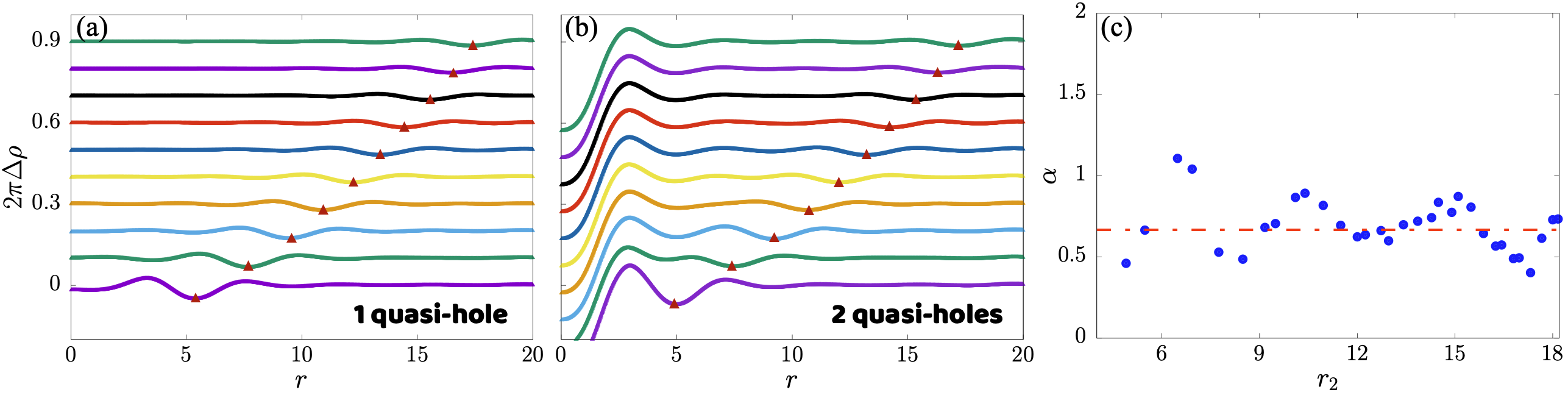}        
        \caption{\label{berry} (a) The density difference between a quasihole in the $m$-th orbital and the ground-state, with a shift applied to the y-axis for convenience, where $m$ ranges from 5 to 50 with an interval of 5. (b) The density difference between two quasiholes in both the $m$-th orbital and the origin and the ground-state. (c) The fractional statistics parameter as varying the location of the second quasihole $r_2$, aligning with the predicted theoretical value. }
    \end{minipage}
\end{figure}
\end{widetext}

Equivalently, one can calculate the mean square radius $\langle r^2 \rangle$ for the electron density, since the radius of the $m$th Landau orbital (with angular momentum $L_z = m\hbar$) is simply $r = \sqrt{2m}l_B$ in the disk geometry~\cite{spin4,Macaluso19,grass20}.
The statistical parameter is given by:
\begin{equation}
	\alpha = (r^2_{\text{1qh}}-r^2_{\text{2qh}})/6
\end{equation}
where the positions of the density minima are represented by $r_{\text{1qh}}$ for the single-quasihole scenario and $r_{\text{2qh}}$ for the two-quasihole scenario. For the FQHE at $\nu = 1/3$, with two independent quasiholes, the predicted statistical parameter is $\alpha = 2/3$~\cite{Jainbook,phase1}. The findings are illustrated in Fig.~\ref{berry}. Panels (a) and (b) illustrate the density difference maps for scenarios involving one and two quasiholes, respectively. Comparison between quasihole state and ground state densities reveals the quasihole position, marked by red triangles. Panel (c) shows the parameter $\alpha$, which corresponds to the theoretical prediction value of $\alpha = 2/3$, highlighting the manifestation of fractional statistics.

\begin{figure}     
\center{\includegraphics[width=8cm]  {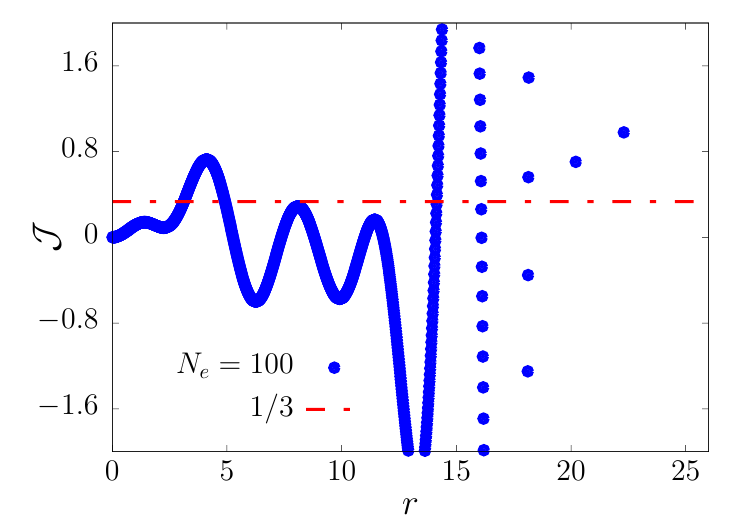}}        
\caption{\label{spin1_3qh} The topological spin of one $e/3$ charged quasihole at the center of disk as formulated in Eq.~(\ref{Jr}). The dotted line is the  theoretically anticipated value $\mathcal{J} = 1/3$. DFT calculation exhibits significantly more pronounced oscillations compared to the MC simulation of the model wavefunction.} 
\end{figure}
\textit{Topological Spin.} Topological spin refers to the quantized angular momentum associated with quasiparticle/quasihole excitations in FQH states. This could be considered a  quantum number associated with the topological phases. Particularly, in disk geometry, it is defined as a measurable quantity linked to the excess angular momentum of a localized quasiparticle relative to the homogeneous bulk state. Mathematically, for a quasiparticle pinned at position $\eta$, the topological spin $\mathcal{J}$
is expressed as\cite{spin1,spin2,spin3,spin4}:
\begin{equation} \label{Jr}
	\mathcal{J}(r) = \int_0^r (\frac{|r'-\eta|^2}{2}-1)(\rho_\eta(r')-\frac{\nu}{2\pi}) d^2r'
\end{equation}
the spin stems from the quadrupole moment characteristic of the charge distribution within the quasiparticle. It adheres to a refined spin-statistics principle, linking it with the topological phase encountered when quasiparticles are exchanged. Notably, it differs from earlier definitions (e.g., on spherical geometries) but remains experimentally accessible through density-profile measurements.  For example, Laughlin quasiholes yield
\begin{eqnarray}
    \mathcal{J} = -\frac{q^2\nu}{2} + \frac{q}{2}
\end{eqnarray} 
where $q$ represents the number of quasiholes, which supports the outcomes of both theoretical predictions and computational simulations. For simplicity, the quasihole is fixed at the center of the disk, namely $\eta=0$, yielding a topological spin $\mathcal{J} \sim 1/3$ as shown in Fig.~\ref{spin1_3qh}. The topological spin calculated via DFT exhibits significantly more pronounced oscillations compared to the Monte Carlo simulations of the model wavefunction presented in Ref.~\onlinecite{spin4}, mainly due to the effects of Coulomb interactions which could also be observed in the electron density as shown in Fig.~\ref{density}.

\begin{widetext}
\begin{figure}[ht]        
    \centering
    \begin{minipage}{\textwidth}
        \centering
        \includegraphics[width=16cm]{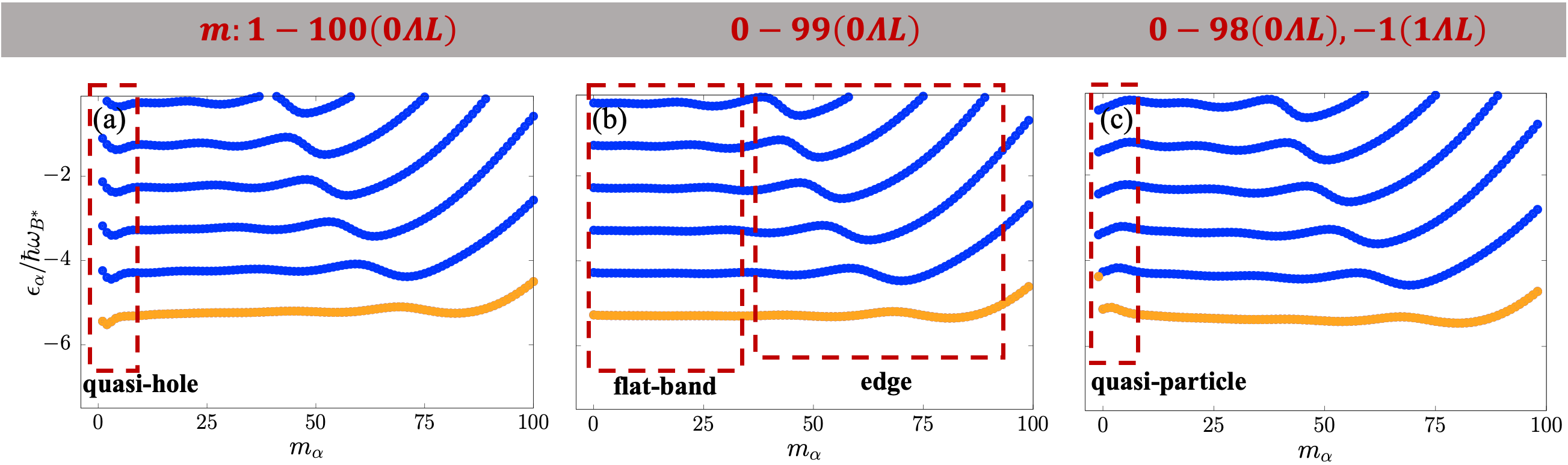}        
        \caption{\label{energypattern} The KS energy $\epsilon_\alpha/(\hbar \omega_{B^*})$ for (a) quasihole, (b) ground, (c) quasiparticle states of filling $\nu=1/3$. The red text of gray area at the top of the figure marks the corresponding filled orbitals. The blue regions represent the orbital energy levels for different $\Lambda$ values, and the yellow regions indicate the actual filled orbital energy levels.}
    \end{minipage}
\end{figure}
\end{widetext}

\textit{Orbital energy pattern.} To further characterize the DFT representation of the FQH ground state and its excitations, the Kohn-Sham orbital energies $\epsilon_\alpha$ are computed through numerical solution of Eq.~(\ref{ksequation1}). Fig.~\ref{energypattern} presents the energy spectra for three characteristic cases at $\nu = 1/3$: (left) the $e/3$ quasihole state, (center) the ground state, and (right) the $e/3$ quasiparticle state. Specifically, the blue lines represent the orbital energies of the higher $\Lambda$-levels, while the yellow lines correspond to the actual CF-filled orbits at this filling. They resemble the Landau level of an electron in a disk configuration but include modifications at the locations of quasiparticles/quasiholes and the system's boundary which demonstrates that the Coulomb interaction is a perturbation in CF theory. For a system containing 100 electrons, the quasihole state corresponds to CF fillings with angular momentum $m = 1$ to $100$ in the $0\Lambda L$. The ground state has CF fillings with $m = 0$ to $99$ in the $0\Lambda L$, while the quasiparticle state is defined by CF fillings for $m = 0$ to $98$ in the $0\Lambda L$ and $m = -1$ in the $1\Lambda L$. Apart from displaying edge features universally, the quasihole and quasiparticle states reveal unique energy distributions at the disk's center. In the quasihole state, the energy of the orbit at the origin is elevated, aiding in the expulsion of CF. In contrast, in the quasiparticle state, the energy of the orbit at the origin, even when situated in $1\Lambda L$, is reduced compared to other orbits at the same level, promoting the attraction of CF. Moreover, the entire bulk region of the ground state maintains a flat band structure analogous to the Landau levels.

\section{Magnetoroton excitation}
\label{sec5}
\begin{figure}[ht]        
\center{\includegraphics[width=8cm]  {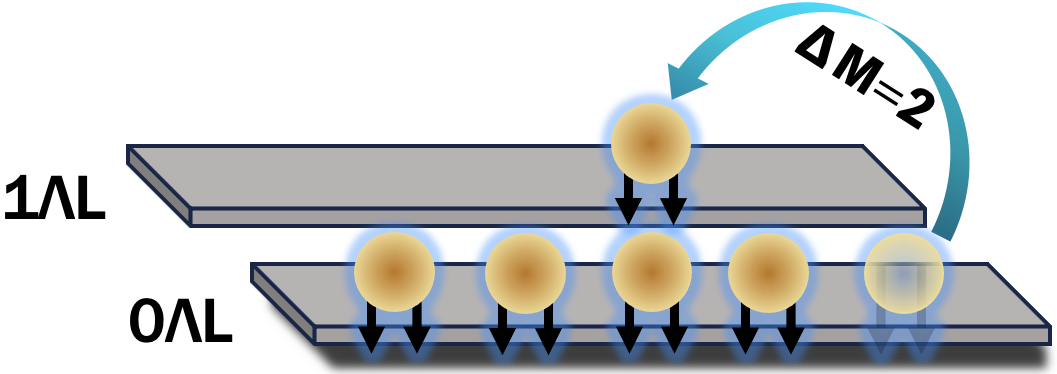}}        
\caption{\label{exciton_model} A schematic cartoon of a single CF exciton for $\nu = 1/3$ FQH state, where the stairs represent the $\Lambda$ levels of CFs. This figure depicts the excitation of a CF from the orbital with the largest angular momentum at the $0$th $\Lambda$ level to the orbital with angular momentum difference $\Delta M=2$ at the $1$st $\Lambda$ level. } 
\end{figure}
Another interesting aspect of FQH systems is their neutral collective excitations, characterized by a magnetoroton mode featuring a distinct roton minimum~\cite{mini1,DTS1,mini2}.  Girvin, MacDonald, and Platzman introduced the single-mode approximation to characterize the lowest-energy neutral excitations as a density wave called the magnetoroton~\cite{GMP2,GMP1}. Wave functions for magnetoroton have been explicitly proposed using multicomponent CF approaches and Jack polynomials~\cite{RodriguezPRB12,YHZZH}. In the past decade, experimentalists have notably advanced in observing these collective modes. Specifically, recent polarization-based Raman scattering experiments have revealed graviton-like excitations and their chirality by means of resonant inelastic scattering peaks~\cite{dulingjie1}. Normally, the topological orders in FQH liquids are examined at the edges because of the bulk-edge correspondence principle. However, various partially understood complexities at the edge complicate the interpretation of edge-related experiments, making them challenging and occasionally ambiguous. As suggested in Ref.~\onlinecite{spectralfunction3,PhysRevResearchSon}, exploring the graviton mode excitations in the bulk can accurately reveal the topological order of FQH liquids, which deserves further research. The majority of numerical investigations into the magnetoroton are conducted on closed manifolds, such as sphere and torus. Its study in disk geometry with an open boundary is still missing though its usefulness is limited due to minimal symmetry and the reduced size of systems that can be accessed.
\begin{figure} [ht]    
\includegraphics[width=8cm]  {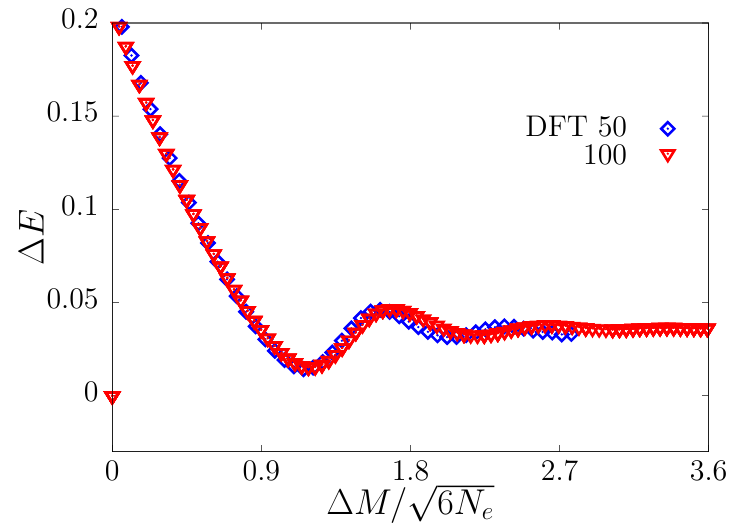}       
\caption{\label{roton} The dispersion of magnetoroton mode for $\nu=1/3$ FQH state with 50 and 100 electrons from DFT. }
\end{figure}

In a recent work~\cite{DFT8}, we developed CF excitons in disk geometry, thereby broadening the exploration of excited states in this geometry. Using both DFT and MC techniques, we analyzed the dispersion of the magnetoroton mode and matched our results with the ED results. Furthermore, our analysis of the spectral function allowed us to determine its spin-2 chirality, affirming the validity of our methodology. We considered a pair of quasihole-quasiparticles composed of CFs, that is, the exciton as shown in Fig.~\ref{exciton_model}, and calculated its energy. Defining the wave vector $k = \Delta M / R_d$, where $\Delta M=M_{\text{gs}}-M_{\text{exciton}}$ is the difference in angular moments between the ground state and the exciton state, and $R_d = \sqrt{6N_e}$ is the radius of the disk. We employ DFT to determine the exciton energy for two systems with varying electron numbers, illustrated in Fig.~\ref{roton}. By setting the horizontal axis to $k$, the data from both systems align on a unified dispersion curve, indicating the consistency between these different sizes. Our findings distinctly showcase a prominent roton minimum and an excitation gap similar to other numerical methods in sphere or torus geometries. In addition, the chirality of the excitation is determined through spectral function calculations, which aligns with prior results obtained through exact diagonalization.

It is important to note that the energy dispersion shown in Fig.~\ref{roton} as derived from DFT exhibits higher energy, particularly near the long-wavelength limit, when compared to results from the Monte Carlo simulations. This discrepancy arises due to the absence of a projection onto the lowest Landau level (LLL) in the DFT method, which is inherently included in MC simulations through wavefunction constraints. As discussed in Ref.~\onlinecite{DFT8}, the projection of the LLL remains a subject for further investigation in this field.

\section{conclusion and prospect}\label{sec6}
This studies have demonstrated that DFT, within the CF framework, efficiently captures both ground state and excitation properties of FQHE. In particular, it reproduces magnetoroton dispersions and topological signatures-fractional charge and statistics-of quasiparticles. In particular, DFT provides a computationally efficient platform for modeling realistic systems with edge potentials and disorder, complementing traditional wavefunction-based methods constrained by idealized geometries. The quantitative agreement of DFT-derived energies and densities with established results from Monte Carlo simulations and experimental observations underscores its reliability in probing FQH physics.  

Despite these successes, challenges remain. The development of accurate exchange-correlation functionals capable of capturing strong correlations, topological order, and fractional statistics remains a critical frontier. Although current approximations work well in specific filling factors, they need improvement to capture the complex interactions such as for higher LLs or the non-Abelian FQH state which is not well defined in CF theory. Additionally, integrating the LLL projection into DFT frameworks is essential to improve accuracy, particularly in the description of low-energy excitations. Comparative studies with experimental data, especially for larger systems and complex geometries, will further validate and refine DFT methodologies. Looking ahead, several promising open questions emerge:

1. LLL-Projected Functionals: Developing functionals that inherently incorporate LLL projection will enhance the precision of DFT in describing FQH states, particularly for low-energy collective modes like the magnetoroton.

2. Time-dependent DFT (TDDFT): Extending DFT to non-equilibrium dynamics using TDDFT holds immense potential for modeling quench protocols, transport phenomena, and time-resolved excitations. We have implemented the TDDFT via fourth-order Runge-Kutta methods~\cite{Yangunpublished}, as demonstrated in the Appendix~\ref{TDDFT}, which provides a blueprint for studying non-equilibrium dynamics in FQH systems.

3. Machine Learning-Augmented Functionals: To investigate more intricate FQH states such as non-Abelian FQHE, or include the experimental device details such as the layer thickness \textit{et al.}, incorporating machine learning methodologies to craft adaptive exchange-correlation functionals specifically suited to the many-body correlations of the FQHE offers a groundbreaking approach.

The collaboration between DFT and novel computational capabilities has already opened new avenues in FQH research, enabling the modeling of larger systems and the investigation of more interesting topological phases and dynamic properties. With the ongoing progress of theoretical and computational methodologies, DFT has contributed to the understanding of the FQH effect and continues to provide new perspectives, complementing other established methods.

\appendix
\section{Time evolution dynamics} \label{TDDFT}
\begin{figure} [ht]     
\center{\includegraphics[width=8cm]  {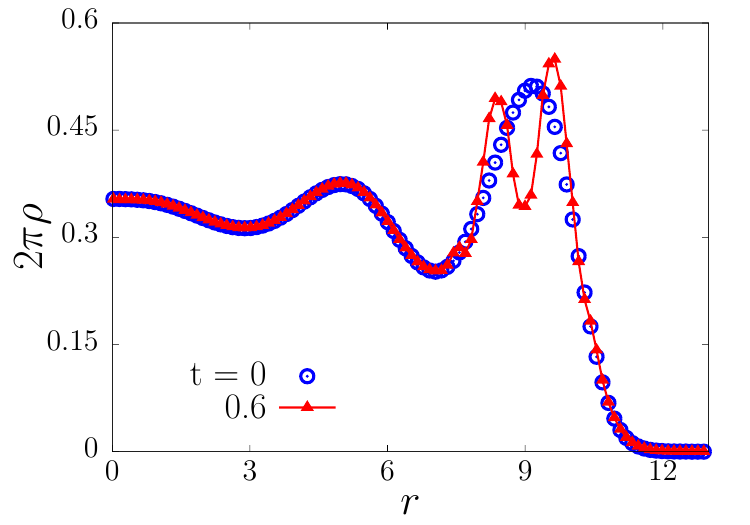}}        
\caption{\label{densitytime} Time evolution of the density $\rho(t)$ in the presence of $V_{\text{disorder}}(r)$.}
\end{figure}
This section outlines our initial findings on the temporal progression of the FQH system, employing the classic fourth-order Runge-Kutta technique for computations over each KS orbital. TDDFT efficiently models non-equilibrium quench dynamics induced by abrupt variations in diverse potential energies. The initial density and the wave function at $t=0$ is defined as $\rho(0)$ and $\psi_\alpha(0)$. Subsequently, the fourth-order Runge-Kutta method is applied to determine the evolution of both the wave function and density at successive time intervals as the system is subjected to the time-dependent KS potential, $V^*_{\text{KS}}(t)$. Considering the time-dependent Schr\"odinger equation of each KS orbital
\begin{equation}
\label{Schodinger}
	\frac{d\psi_\alpha(t)}{dt}=f(t,\psi_\alpha)=-\frac{\mathrm{i}}{\hbar}H(t)\psi_\alpha(t)
\end{equation}
where $H(t)=T^*+V^*_{\text{KS}}(t)$, Eq.~(\ref{Schodinger}) is discretized and iteratively computed. Assuming that time is discretized as $t_n$, with a time step $h$, i.e., $t_{n+1}=t_{n}+h$, the next step of the solution is given by
\begin{equation}
	\psi_{n+1}=\psi_n+\frac{h}{6}(k_1+2k_2+2k_3+k_4)
\end{equation}
where the coefficients are defined as
\begin{eqnarray}
    	k_1&=&f(t_n,\psi_n)=-\frac{\mathrm{i}}{\hbar}H \cdot \psi_n, \nonumber\\
		k_2&=&f(t_n+\frac{h}{2},\psi_n+\frac{h}{2}k_1)=-\frac{\mathrm{i}}{\hbar}H\cdot(\psi_n+\frac{h}{2}k_1), \nonumber\\
		k_3&=&f(t_n+\frac{h}{2},\psi_n+\frac{h}{2}k_2)=-\frac{\mathrm{i}}{\hbar}H\cdot(\psi_n+\frac{h}{2}k_2), \nonumber\\
		k_4&=&f(t_n+h,\psi_n+h k_3)=-\frac{\mathrm{i}}{\hbar}H\cdot(\psi_n+h k_3).
\end{eqnarray}
 Starting from the initial wave function, the evolution of the wave function at each time step is computed iteratively using the above procedure. This method allows us to explore the system's evolution under dynamic perturbations, with a particular focus on observing changes in topological properties and the potential for dynamical phase transitions. Such studies provide valuable insights into the time-domain response of complex quantum systems and their behavior under interactions with external drives.
 As a concrete example, the system's response to an impurity potential $V_{\text{disorder}}$ introduced at $t>0$ can be examined. The analysis begins with the potential generated by a uniformly distributed charge:
\begin{equation}
	\begin{split}
		V(r) & = \int \frac{\sigma dr'^2}{|r-r'|} \\
		&= \frac{2}{\pi} \int_0^R \frac{\sigma 2\pi r'dr'}{|r+r'|}  K[\frac{4r\cdot r'}{(r+r')^2}]  \\
	\end{split}
\end{equation}
where $K(x)=\int_0^{\pi/2} \frac{d\theta}{\sqrt{1-x\sin^2(\theta)}}$ is the complete elliptic integral of the first kind with $0\leq x\leq 1$. This represents the electric potential due to a uniformly charged disk with charge density $\sigma$ distributed across $r'=0-R$ likewise the background confinement potential. The electrostatic potential from a thin circular ring at radial position $r'$ takes the form:
\begin{equation}
	\begin{split}
		V_{\text{disorder}}(r) &= \frac{2}{\pi} \frac{\sigma 2\pi r' \Delta r'}{|r+r'|}  K[\frac{4r\cdot r'}{(r+r')^2}]  \\
		&=\frac{2Q}{\pi |r+r'|}  K[\frac{4r\cdot r'}{(r+r')^2}]  \\
	\end{split}
\end{equation}
the analysis employs a ring potential with parameters $Q=10, r'=9$, representing a localized ring potential of strength $Q$ at radial distance $r'$ on the disk.
Fig.~\ref{densitytime} presents the time evolution of the system after the ring potential is introduced. The charge distribution exhibits a significant decrease in electron density at $r'$, creating a prominent dip that warrants more in-depth analysis of the dynamics.

\acknowledgments
This work was supported by National Natural Science Foundation of China Grants No. 12474140 and 12347101. Y. Yang was supported by graduate research and innovation foundation of Chongqing, China Grant No. CYB25066 and the inaugural Doctoral Student Special Project of the China Association for Science and Technology Young Talents Lifting Program (2024). Yayun Hu was supported by National Natural Science Foundation of China Grants No. 12204432.

\bibliography{biblio_fqhe.bib}

\end{document}